\documentclass[12pt, preprint]{aastex}

\begin{document}

\title{A Study of the Unusual Z Cam Systems IW Andromedae
and V513 Cassiopeia\altaffilmark{*}}

\author{Paula Szkody\altaffilmark{1,2},
Meagan Albright\altaffilmark{1},
Albert P. Linnell\altaffilmark{1},
Mark E. Everett\altaffilmark{3},
Russet McMillan\altaffilmark{4},
Gabrelle Saurage\altaffilmark{4},
Joseph Huehnerhoff\altaffilmark{4},
Steve B. Howell\altaffilmark{5,2},
Mike Simonsen\altaffilmark{6},
Nick Hunt-Walker\altaffilmark{1}}

\altaffiltext{*}{Based on observations obtained with the Apache Point
  Observatory (APO) 3.5-meter telescope, which is owned and operated
  by the Astrophysical Research Consortium (ARC).}
\altaffiltext{1}{Department of Astronomy, University of Washington,
  Box 351580, Seattle, WA 98195; 
szkody@astro.washington.edu}
\altaffiltext{2}{Visiting Astronomer, Kitt Peak National Observatory, National
Optical Astronomy Observatory, which is operated by the Association of Universities for Reasearch in Astronomy (AURA) under cooperative agreement with the National Science Foundation}
\altaffiltext{3}{National Optical Astronomy Observatories, 950 N. Cherry Ave, Tucson, AZ 85719}
\altaffiltext{4}{Apache Point Observatory, Sunspot NM 88349}
\altaffiltext{5}{NASA Ames Research Center, Moffett Field, CA 94035}
\altaffiltext{6}{49 Bay State Road, Cambridge, MA 02138; mikesimonsen@aavso.org}
\altaffiltext{2}{Visiting Astronomer, Kitt Peak National Observatory, National
Optical Astronomy Observatory, which is operated by the Association of Universities 
for Reasearch in Astronomy (AURA) under cooperative agreement with the National 
Science Foundation}

\begin{abstract}
The Z Cam stars IW And and V513 Cas are unusual in having outbursts following 
their standstills in contrast to the usual Z Cam behavior of quiescence 
following standstills. In order to gain further understanding of these 
little-studied systems, we obtained spectra correlated
with photometry from the AAVSO throughout a 3-4 month interval in 2011. In 
addition, time-resolved spectra were obtained in 2012 that provided orbital 
periods of 3.7 hrs for IW And and 5.2 hrs for V513 Cas. The photometry of V513 
Cas revealed a regular pattern of standstills and outbursts with little time at quiescence, while IW And underwent many excursions from quiescence to outburst 
to short standstills. The spectra of IW And are similar to normal dwarf novae, 
with strong Balmer emission at quiescence and
absorption at outburst. In contrast, V513 Cas shows a much flatter/redder 
spectrum near outburst with strong HeII emission and prominent emission cores 
in the Balmer lines. Part of this continuum difference may be due to reddening 
effects. While our attempts to model the outburst and standstill states of IW 
And indicate
 a mass accretion rate near 3$\times$10$^{-9}$M$_{\odot}$yr$^{-1}$, we could
find no obvious reason why these systems behave differently following 
standstill compared to normal Z Cam stars.
\end{abstract}


\section{Introduction}

Cataclysmic variables (CVs) are close binary systems in which a cool secondary
star fills its Roche lobe and transfers mass to a white dwarf primary
(Warner 1995). The CVs that undergo outbursts related to their mass transfer
rate and subsequent accretion rate onto the white dwarf are termed dwarf novae.
The Z Cam systems are a subgroup 
which generally have relatively long orbital periods (3-5 hrs), relatively
high mass transfer and accretion rates and relatively low amplitude (3 mag) outbursts
compared to the majority of dwarf novae. The major identifying trait of
the Z Cam systems is that they show standstills in their light curves when
they remain at about 0.7 magnitude below their outburst brightness for
weeks to months. A number of review articles describe a
 model for dwarf novae outbursts and an explanation for the Z Cam phenomena; 
e.g. Smak (1984), Cannizzo (1993), Osaki (1996),
Lasota (2001), Buat-M\'{e}nard et al. (2001a,b). 

The basic premise behind standstills is that there
is a critical rate of mass accretion that determines outburst behavior.
This critical rate is model dependent but depends on the parameters of
disk radius, viscosity, orbital period
and mass ratio of the system. 
The systems with rates below the critical value 
(estimated to be $<$3$\times$10$^{-9}$M$_{\odot}$yr$^{-1}$ by Buat-M\'{e}nard et
al. 2001b for Z Cam) have dwarf nova
outbursts while those above the critical rate are novalike systems that 
remain in a constant high state. The Z Cam systems are thought to be very
close to the critical rate. An increase in mass transfer can heat the outer edge of the
accretion disk and decrease the critical accretion value so that the disk stays in
a stable high state.  By taking into account the 
mass transfer stream-impact on the disk and tidal torques, 
Buat-M\'{e}nard et al. (2001b) 
were able to reproduce standstills with mass transfer variations of
a factor of two for Z Cam.
Hartley et al. (2005) compared FUSE observations at
quiescence with ORFEUS standstill data on Z Cam itself and showed that
the standstill accretion rate of 1$\times$10$^{-9}$M$_{\odot}$ yr$^{-1}$ was consistent 
with a lower critical value at standstill than at outburst.

Recently, Simonsen (2011) coordinated an observing campaign (Z CamPaign) that 
utilized
the AAVSO to correctly classify Z Cam systems and monitor their long term
light curves. He identified two systems with peculiar behavior following
a standstill, IW And and V513 Cas. While the majority of Z Cam systems decline 
to quiescence
after a standstill, these systems usually showed a rise to outburst after
standstill. In an attempt to understand this peculiarity and determine the 
characteristics of these
two systems, we obtained spectra at different outburst states throughout
a 3-4 month interval in 2011 as well as time-resolved spectra in 2012 
to determine the orbital
periods of these little-studied systems. We also made an attempt to model
the spectra of IW And from quiescence to outburst to find the corresponding
accretion rates in comparison to those for Z Cam.

\section{Observations}

The spectra were obtained at Kitt Peak National Observatory (KPNO) and at
the Apache Point Observatory (APO). At KPNO, the RC-Spectrograph was used
on the Mayall 4m telescope with the 2048 CCD T2KA and a 1 arcsec slit.
Grating KPC-22b in second order enabled a spectrum with good focus from
3800-4900\AA\ with a resolution of 0.7\AA\ pixel$^{-1}$. Standard stars as
well as FeAr lamps
were taken for flux and wavelength calibrations. At the 3.5m APO telescope, 
the Double-Imaging
Spectrograph was used in either low (L) or high (H) resolution mode to provide
simultaneous blue and red spectral coverage. The low resolution gratings
provided useful coverage from 3400-9500\AA\ with a resolution of 1.2\AA\ 
pixel$^{-1}$ in the blue and 2.3\AA\ pixel$^{-1}$ in the red. The high
resolution gratings enabled a resolution of 0.6\AA\ pixel$^{-1}$ over
wavelengths of 3900-5000\AA\ in the blue and 6000-7200\AA\ in the red.
Flux standards and HeNeAr lamps were used for calibration.

Tables 1 and 2 summarize the observations that were obtained.

For both datasets, IRAF\footnote{IRAF is distributed by the National 
Optical Astronomy Observatory,
which is operated by the Association of Universities for Research in Astronomy, 
Inc.,
under cooperative agreement with the National Science Foundation.}
routines were used to accomplish bias and flat field corrections and to
transpose the observed data to wavelength and flux. For the high
resolution time-resolved spectra obtained during 2012 October, velocity
curves were obtained by measuring the 
emission lines of H$\alpha$ and H$\beta$ with the centroid ${\it e}$ routine in
${\it splot}$ and using an IDL program to find the best fit of the
velocities to a sine-curve.

For each observation, the state of the system was obtained from the
American Association of Variable Star Observers\footnote{http://www.aavso.org}
archive and included in Tables 1 and 2. The V magnitudes that are available 
for each system during 2011 are plotted in Figure 1. 

\section{Light Curves}

The light curves plotted in Figure 1 show the quiescent, outburst and standstill
states of each of the two objects. V513 Cas has better coverage and shows
four well-defined standstills. In each case, the standstill is followed by
a rise to outburst rather than by a decline to quiescence. Unfortunately,
the time spent at quiescence is very short and comparison with Table 1 shows
that the standstills are well-covered but 
none of the spectroscopic times coincided with a quiescent state. The 
time-resolved spectra obtained in 2012 were obtained during the peak of
an outburst that followed months at standstill.

IW And (Figure 1) has sparser coverage near the start of the spectroscopic
observations. This system has many more excursions to quiescence and Table
2 shows that spectra were obtained at all states. There are two possible
standstills, a short one near days 780-790 and a longer one during
days 820-840. In each case, the standstill resulted in a rise to outburst.
The 2012 October 29 time-resolved spectra were obtained at an interesting time. During the course of the 4.75 hrs of observation, IW And was transitioning
to a standstill state following a decline after a brief outburst on October
23 (available AAVSO measurements are V=15.0 on 26 October and V=14.4 on 30 
October).

While both objects have roughly 3 mag differences between quiescence and
outburst (IW And is 3.4 while V513 Cas is 2.8), and the usual 0.7 mag 
difference between outburst and standstill,
the longer and more frequent standstills apparent in V513 Cas imply its
mean mass accretion rate is
closer to its critical $\dot{M}$ value.

\section{Spectral States}

The data acquired in 2011 June over 4-6 consecutive nights (JD2455722-2455727) 
document the
changes that occur from quiescence to outburst in IW And (Figure 2)
and through a decline from outburst in V513 Cas (Figure 3). IW And
shows the usual spectral changes of dwarf novae from Balmer emission
lines at quiescence to deep absorption near outburst as the disk
tranforms to the hot outburst state. During the rise on June 13, the
lag in the rise of the blue light compared to the red is very evident. This
lag has been well documented from many past studies of dwarf novae, 
and is evident in
both outside-in and inside-out outbursts (Buat-M\'{e}nard et al. 2001a).
During the last 2 hrs of the 25 spectra obtained on 2012 October 29,
the Balmer lines show a rapid transition from emission near quiescence
to a standstill configuration with broad absorption flanking strong emission 
cores (Figure 4). In this rapid rise, the blue continuum flux also seems to lag
in the rise.

In contrast to this normal behavior, the spectra of V513 Cas are unusual
in the lack of a strong blue continuum near outburst and the weakness of
the Balmer emission lines during the decline to quiescence (Figure 3).
In addition, the HeII 4686 line is very strongly in emission on the decline
from outburst
on June 9. As the galactic latitude of V513 Cas is only +3 deg, it is likely
that reddening can account for some part of the lack of blue continuum.
Since the distance is not known and there is no quiescent spectrum to reveal
lines from the secondary star, it is not possible to obtain a value
for the reddening. The APO spectra obtained at the peak of outburst on
2012 October 6 (Figure 5) are very similar to the 2011 June 9 spectrum, with the
strong HeII line, the emission core in H$\beta$ and a similar rising
slope to the red wavelengths. At standstill (Figure 5), the HeII line
is still present, although the slope of the continuum flattens out. The
standstill spectrum of IW And is also compared to its outburst in Figure 5. 
Its outburst spectrum is similar but has more blue flux than at standstill.

\section{Orbital Period Determination}

In order to compare the properties of the two systems in more detail, it
was necessary to determine the orbital periods, which relate to the
accretion rates and especially the critical rates. 
Since the 23 time-resolved spectra of V513 Cas were
obtained at the peak of outburst and the 25 time-resolved spectra of
IW And during a rise to standstill, the blue spectra are difficult
to measure due to the broad absorption lines from the accretion disk.
However, the red spectra have H$\alpha$ in emission which could be measured
and result in the radial velocity curves shown in Figure 6. 
As many of the IW And
blue spectra showed emission lines of H$\beta$ before the standstill state was
reached, this line was also measured. The parameters for the least-squares fit 
of the velocities to a sine-wave
are listed in Table 3, where the $\sigma$ is the standard deviation of the
entire fit. For V513 Cas, the 4.3 hrs of spectral coverage was
best fit with a period of 312 min (5.2 hrs) while the 4.7 hrs of coverage
on IW And yielded a period of 223 min (3.7 hrs). 
Due to the limited length of the spectral
coverage, especially for V513 Cas, these orbital periods should only be used
as estimates. However, the values place
IW And near the lower limit of Z Cam stars and V513 Cas in the middle
(Table 3.2 of Warner 1995). The longer orbital period of V513 Cas implies
that its mean mass transfer/accretion rate and $\dot{M}_{crit}$ are larger than IW And. 

\section{Modeling IW And}

Given the uncertainty in the reddening of V513 Cas and the lower signal-to-noise
of its spectra, we chose IW And and the sequence of spectra obtained in 2011 
June, as well as the standstill spectra on 29 September, to try to
approximate a model for its accretion disk at quiescence, rise, near outburst
and standstill.
Table 4 contains a list of our adopted system parameters. For consistency
with the tables of Knigge (2006) and Knigge et al. (2011), we adopted a
white dwarf mass of 0.75M$_{\odot}$. 
With our observed orbital period of 223 min, Figure 16 of Knigge et l. (2011) provides
a T$_{eff}$ of 25,000K for the white dwarf. Using this temperature, the radius 
of a white dwarf with a carbon core and an overlying helium layer is 
0.015R$_{\odot}$, (Figure 4a 
of Panei et al. 2000). A secondary mass of 0.27M$_{\odot}$ 
is obtained from Table 3 of Knigge (2006), and results in a mass ratio q of 
0.36 for the system. Howell et al. (2001) found that secondary stars with periods
longer than 3 hrs have bloated secondaries, which would increase the secondary
radius and q values slightly, but the secondary has little effect on the outburst
and standstill spectra. The observed radial velocity amplitude constrains
the orbital inclination to about 30 degrees. The M$_{V}$-P$_{orb}$ relations 
from Harrison et al. (2004) and Patterson (2011) provide a distance of about 
700 pc that is used as a scaling factor between the models and the observed
spectra.
 
Figure~7 presents a projection of the system on the plane of the sky while 
Figure~8 shows the model radial velocity curves of the two stellar components 
(the observed velocities shown in Figure 6 
are from the emission lines originating in the 
accretion disk).
 
We do not have a time-dependent outburst model available to calculate
theoretical synthetic spectra as a function of time during an outburst cycle;
consequently, we model conditions at outburst and standstill with hot
stable disks, with mass accretion rate equal to the mass transfer rate. 
We therefore used BINSYN (Linnell et al. 2012),
 which calculates models of stable 
cataclysmic variables in the hot state, with input annuli calculated with 
TLUSTY (Hubeny 1988), and SYNSPEC (Hubeny et al. 1994) to model the
observed outburst spectrum of 2011 June 14.
Under normal operation, BINSYN is supplied with a set of annulus synthetic 
spectra for a specified mass transfer rate and white dwarf mass. BINSYN then 
uses the specified mass transfer rate and range of radii in the accretion disk 
to calculate theoretical T$_{eff}$ values as a function of annulus radius. If 
there are many more narrow width annuli specified than available synthetic 
spectra, BINSYN interpolates to the specified annuli  T$_{eff}$ values. In addition to 
calculating a synthetic spectrum for
the accretion disk face, BINSYN also calculates a synthetic spectrum for the 
accretion disk rim and for the white dwarf and the secondary star 
(see Linnell \& Hubeny 1996, for details).  

We used TLUSTY and SYNSPEC to calculate complete steady state accretion discs 
for mass transfer/accretion
 rates of 1.0, 2.0, and 3.0 $\times$10$^{-9}$ M$_{\odot}$yr$^{-1}$ with a 
viscosity parameter $\alpha$ = 0.10.
In each case we represented the accretion disk with 40 annuli in two groups of 
equal width annuli. The first group of 10 annuli extended from the white
dwarf equator to 
the radius of maximum T$_{eff}$ at 1.36 R$_{wd}$. The second group extended 
from the first group boundary to the cutoff radius specified by BINSYN.  
The synthetic spectrum fits to the observational data at outburst are sensitive 
to the mass transfer rate. Both the rates of 1.0 and 2.0$\times$10$^{-9}$
 had too little flux in the 
3800-4000\AA\ region while the value of 3$\times$10$^{-9}$ provided the best 
fit. With the distance specified, the outer radius of the accretion disk was 
modified by trial and error until the system 
output synthetic spectrum from BINSYN fitted the outburst spectrum as closely 
as possible. The outer radius of the standard model accretion disk thus 
determined was 0.14R$_{\odot}$ (1.0$\times$10$^{10}$ cm for the outburst model).
The fit to the outburst spectrum is shown in Figure 9 and this disk size is
shown in Figure 7.

If we assume the same steady mass transfer/accretion rate exists at all states, the
flatter slopes of the rising to outburst and quiescent observed spectra
 (data from 2011 June 13 and 11 respectively) point to a lower excitation
accretion disk. To fit both the continuum and the depths of the Balmer lines
in the rising spectrum, a source with T$_{eff}$ of
$\sim$13,000K is needed. Consequently, we successively removed the inner
highest temperature annuli input
to BINSYN until a fit was achieved. This led to the fit shown in Figure 10,
with the highest T$_{eff}$ annulus at 12,479K at a distance of 8R$_{wd}$ and
an outer disk radius of
0.27R$_{\odot}$ (1.88$\times$10$^{10}$ cm). 
While this fit looks quite good, the nature of
the disk during the rise to outburst is likely not in a a steady state. However,
the temperature fits give some measure of the conditions of the disk material
during that time.

For the quiescent spectrum, we note the presence of Balmer emission lines suggesting
the possibility of an optically thin region of the accretion disk (see the 
comments by Patterson 2011). On the other hand, Idan et al. (2010) find
that quiescent accretion disks are optically thick with a T$_{eff}\sim$5000K. 
We supplied BINSYN with solar composition stellar synthetic spectra for T$_{eff}$ values of 3500K and 7000K for interpolation, and used the BINSYN option to 
specify explicit
T$_{eff}$ values for accretion disk annuli; we specified 5000K for the 
outermost 32
 annuli and 6000K for the innermost 8 annuli. The reason for the slight 
temperature gradient was to produce a slightly flatter continuum for comparison with the
observed spectrum. Since the annulus T$_{eff}$ values are specified, our result does
 not depend on a particular viscosity parameter value. The quiescent continuum 
source cannot have a
T$_{eff}$ much above 5000K without producing a discrepancy with the observed continuum
 slope.
The empirical procedure of adjusting the outer radius of the accretion disk to 
produce a fit to the observed quiescent spectrum, using the distance scaling factor, 
led to an outer radius of the accretion disk of 0.08R$_{\odot}$. Figure 11
shows this fit along with the contributions of the white dwarf and the
secondary star to the quiescent spectrum. The dominance of the accretion
disk over that of the underlying stars is confirmed by the lack of any
TiO bands or rise in flux evident in the red spectrum obtained at
quiescence on 2011 July 5.

Applying a similar method to the standstill spectra from 2011 September 29,
using the same set of annuli as the rising to outburst model results in
a radius of the disk of 0.33R$_{\odot}$. This fit is shown in Figure 12.
The model Balmer lines are deeper than observed but the observed lines
are filled in by emission cores.

We can compare our values to the best studied system, Z Cam. Hartley et al.
(2005) used ultraviolet data from FUSE and ORFEUS to obtain model
parameters for the system at quiescence and standstill. They found a
white dwarf temperature near 57,000K and a disk at standstill accreting
at about 1$\times$10$^{-9}$M$_{\odot}$ yr$^{-1}$, which is about a factor
of 4 below the accretion rate at outburst (Knigge et al. 1997). The 
orbital period of Z Cam (7 hrs) is about twice that of IW And so the
expectation of a hotter white dwarf and a higher mass accretion rate
compared to IW And is reasonable. Since the standstill disk accretion rate
was shown to be less than outburst, we attempted to also fit IW And with
a reduced rate. Using a mass accretion rate of 1.0$\times$10$^{-9}$M$_{\odot}$
yr$^{-1}$ and the hot radii extending to the white dwarf equator results
in an outer radius of the accretion disk of 0.16R$_{\odot}$. This model fit
is shown in Figure 13. The high order Balmer absorption lines are fit
better with this lower accretion rate, but the continuum falls off too
steeply to provide a good fit at longer wavelengths. Thus, neither of
our two models represent the standstill situation well. Given the short
wavelength range of our optical data, and the lack of ultraviolet
spectra to determine the actual white dwarf  T$_{eff}$, further refinements
must await more extensive coverage.

\section{Conclusions}

Our photometric and spectroscopic monitoring program over 3-4 months has
revealed some new information on the little-studied systems of IW And
and V513 Cas. IW And is a very variable system, showing constant excursions
from quiescence to outburst, with infrequent standstills. In contrast,
V513 Cas spends about half its time in a standstill or outburst state, with
only brief returns to quiescence about once per month. The spectra of
IW And at different states are very typical of dwarf novae, with strong
Balmer emission at quiescence and deep Balmer absorption at outburst. 
While the quiescent state of V513 Cas was not captured, the outburst
and standstill states are flat or decreasing to the blue, while the high
excitation line of HeII is strongly in emission. Our time-resolved
spectra provide orbital periods near 3.7 hrs for IW And and 5.2 hrs for
V513 Cas.

Our modeling efforts on the outburst of IW And indicate a mass accretion
 rate of 3$\times$10$^{-9}$M$_{\odot}$ yr$^{-1}$ for a white dwarf mass of
0.75M$_{\odot}$ and corresponding parameters from relations in the literature
for the secondary (0.27M$_{\odot}$) and a distance of 700 pc. This same rate of 
mass accretion can fit the rise
to outburst and the quiescent spectra if the radius of the disk
is altered. The fit to the standstill spectra are not as satisfactory,
producing deeper absorption lines than observed. If the accretion rate
is lowered by a factor of 3, the continuum is too steep as compared to
the observations. Further wider wavelength coverage, expecially in
the ultraviolet is needed to formulate a better model for the disk. 
With the data in hand, we could find no obvious reason why the emergence
from standstill to an outburst rather than to quiescence occurs in
these two systems contrary to the normal Z Cam system. Characterising
the basic parameters is just a first step toward a solution to that dilemma.

\acknowledgments
We acknowledge with thanks the variable star observations from the AAVSO 
International Database contributed by observers worldwide and used in this 
research. We are also grateful to Thomas Harrison for obtaining a spectrum 
as part of this program.
This work was partially supported by NSF grant AST-1008734.

{\it{Facility:} \facility{AAVSO}, \facility{Mayall(RC-Spec)}, \facility{APO(DIS)}}

\clearpage
\begin{deluxetable}{lcllccl}
\tablewidth{0pt}
\tablecaption{V513 Cas Spectroscopic Observations}
\tablehead{
\colhead{UT Date} & \colhead{JD} & \colhead{Obs} &{Exp(min)} & \colhead{UT Start} 
& \colhead{V} & \colhead{State}}
\startdata
{05/14/2011} & 2455695.5 & {KPNO} & 10.4 & 11:19:09 & 15.8 & {standstill}\\
{05/15/2011} & 2455696.5 & {KPNO} & 15x2 & 10:55:30 &  15.8 & {standstill}\\
{05/16/2011} & 2455697.5 & {KPNO} & 10 & 11:12:42 & 15.8 & {standstill}\\
{06/09/2011} & 2455721.5 & {KPNO} & 15 & 11:01:55 & 15.4 & {decline}\\
{06/10/2011} & 2455722.5 & {KPNO} & 13.3 & 11:04:41 & - & {decline}\\
{06/11/2011} & 2455723.5 & {KPNO} & 20 & 10:31:45 & - & {decline}\\
{06/12/2011} & 2455724.5 & {KPNO} & 15 & 10:41:09 & - & {decline}\\
{06/13/2011} & 2455725.5 & {KPNO} & 15 & 10:51:59 & 16.4 & {decline}\\
{06/14/2011} & 2455726.5 & {KPNO} & 20 & 10:41:23 & 16.7 & {decline}\\
{07/05/2011} & 2455747.5 & {APO,H } & 15 & 09:36:33 & 15.7 & {standstill}\\
{08/14/2011} & 2455787.5 & {APO,L } & 15 & 05:38:24 & 15.8 & {standstill} \\
{08/20/2011} & 2455793.5 & {APO,L } & 6.7x3 & 07:18:32 & 15.8 & {standstill}\\
{08/26/2011} & 2455799.5 & {KPNO} & 20 & 10:43:24 & 15.8 & {standstill}\\
{08/27/2011} & 2455800.5 & {APO,L} & 6.7x2 & 07:22:23 & 15.8 & {standstill}\\
{09/04/2011} & 2455808.5 & {KPNO} & 16.7 & 10:57:300 & 15.6 & {rise}\\
{09/16/2011} & 2455820.5 & {APO,L } & 10 & 06:58:29 & 15.8 & {decline}\\
{09/29/2011} & 2455833.5 & {APO,H } & 15x2 & 07:35:38 & 15.5 & {decline}\\
{10/06/2012} & 2456206.5 & {APO,H } & 10x23 & 02:37:32 & 15.2 & {outburst}\\
\enddata

\end{deluxetable}
\clearpage

\begin{deluxetable}{lccccclcl}
\tablewidth{0pt}
\tablecaption{IW And Spectroscopic Observations}
\tablehead{
\colhead{UT Date} & \colhead{JD} & \colhead{Obs} &{Exp(min)} &
\colhead{UT Start} & \colhead{V} & \colhead{State}}
\startdata
{06/11/2011} & 2455723.5 & {KPNO} & 20 & 10:55:37 & - & {quiescence} \\ 
{06/12/2011} & 2455724.5 & {KPNO} & 15 & 11:00:01  & - & {quiescence}\\ 
{06/13/2011} & 2455724.5 & {KPNO} & 15 & 11:10:50 & - & {rise}\\ 
{06/14/2011} & 2455726.5 & {KPNO} & 15 & 11:05:31  & -& {outburst} \\
{07/05/2011} & 2455747.5 & {APO,H} & 10 & 09:54:38 & 17.2 & {quiescence}\\ 
{08/14/2011} & 2455787.5 & {APO,L} & 15 & 06:27:25 & 15.0 & {standstill}\\
{08/20/2011} & 2455793.5 & {APO,L} & 5x3 & 07:45:26 & 15.0 & {standstill}\\
{08/26/2011} & 2455799.5 & {KPNO} & 20 & 10:19:08 & 14.0 & {outburst}\\ 
{08/27/2011} & 2455800.5 & {APO,L} & 5x2 & 07:45:45 & 14.0 & {outburst}\\
{09/04/2011} & 2455808.5 & {KPNO} & 20 & 10:01:18 & 17.0 & {quiescence}\\ 
{09/16/2011} & 2455820.5 & {APO,L} & 5 & 06:48:43 & 14.7 & {standstill}\\ 
{09/29/2011} & 2455833.5 & {APO,H} & 10x2 & 07:13:11 & 14.5 & {standstill}\\
{10/29/2012} & 2456229.5 & {APO,H} & 10x25 & 02:03:38 & 14.4 & {standstill}\\
\enddata
\end{deluxetable}

\clearpage
\begin{deluxetable}{lccccc}
\tablewidth{0pt}
\tablecaption{Radial Velocity Solutions}
\tablehead{
\colhead{Object} & \colhead{Line} & \colhead{$\gamma$(km s$^{-1}$)} & 
\colhead{K (km s$^{-1}$)} & \colhead{P (min)} & \colhead{$\sigma$(km s$^{-1}$)} }
\startdata
IW And & H$\alpha$ & -13.8$\pm$0.6 & 46.2$\pm$3.5 & 223 & 11 \\
IW And & H$\beta$ & -61.9$\pm$1.1 & 57.5$\pm$6.5 & 221 & 21 \\
V513 Cas & H$\alpha$ & -35.9$\pm$0.4 & 42.4$\pm$1.9 & 312 & 7 \\
\enddata
\end{deluxetable}

\clearpage
\begin{deluxetable}{lc}
\tablewidth{0pt}
\tablecaption{IW And System Parameters}
\tablehead{
\colhead{Parameter} & \colhead{Value} }
\startdata
Period & 223 min \\
WD T$_{eff}$ & 25,000K \\
M$_{wd}$ & 0.75M$_{\odot}$ \\
R$_{wd}$ & 0.015R$_{\odot}$ \\
M$_{sec}$ & 0.27M$_{\odot}$ \\
q & 0.36 \\
d & 700 pc \\
i & 30$^{\circ}$ \\
\enddata
\end{deluxetable}

\clearpage
\begin{figure}
\plotone{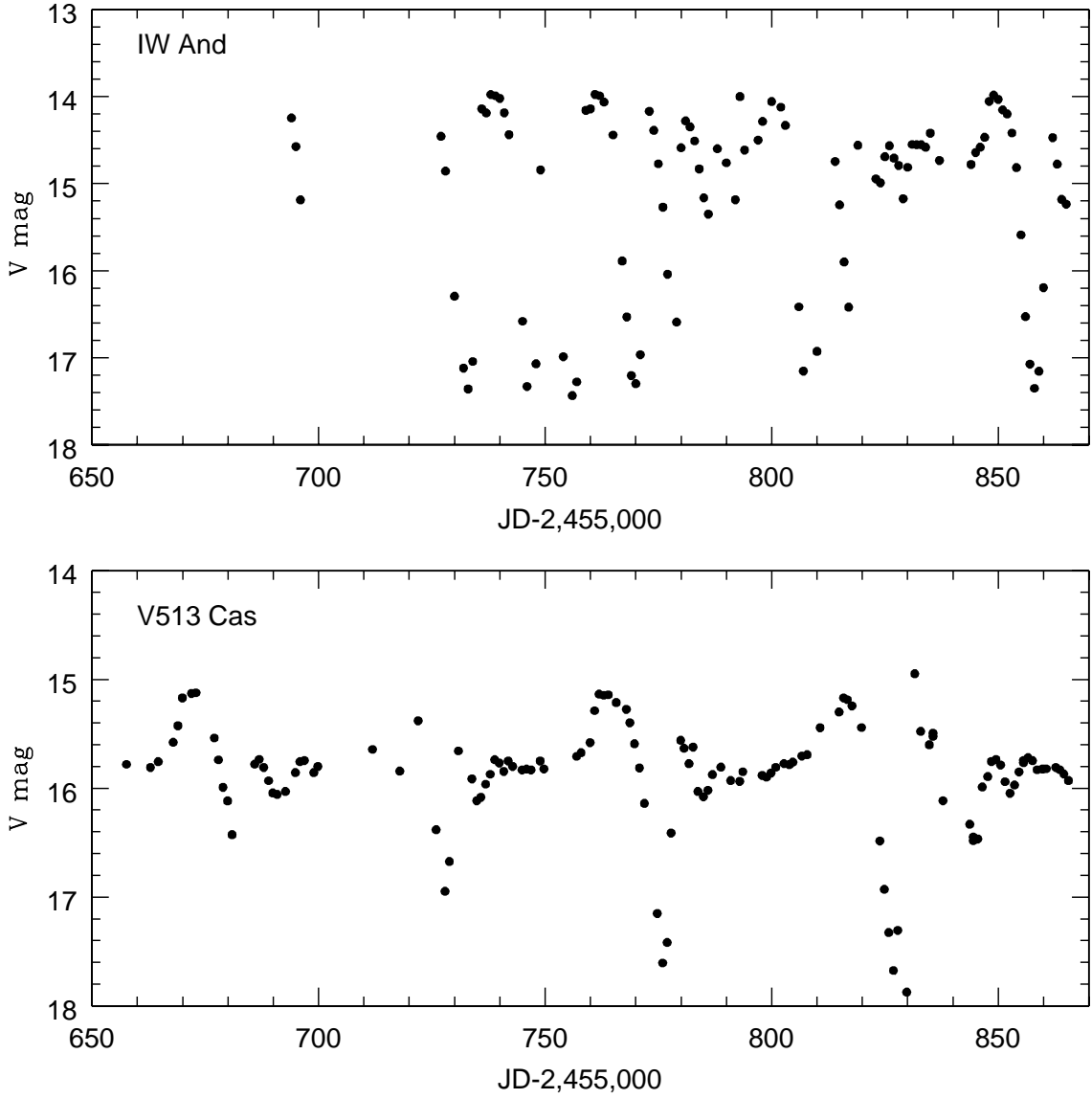}
\caption{AAVSO data during the timespan of our 2011 spectroscopic coverage.}
\end{figure}

\begin{figure}
\plotone{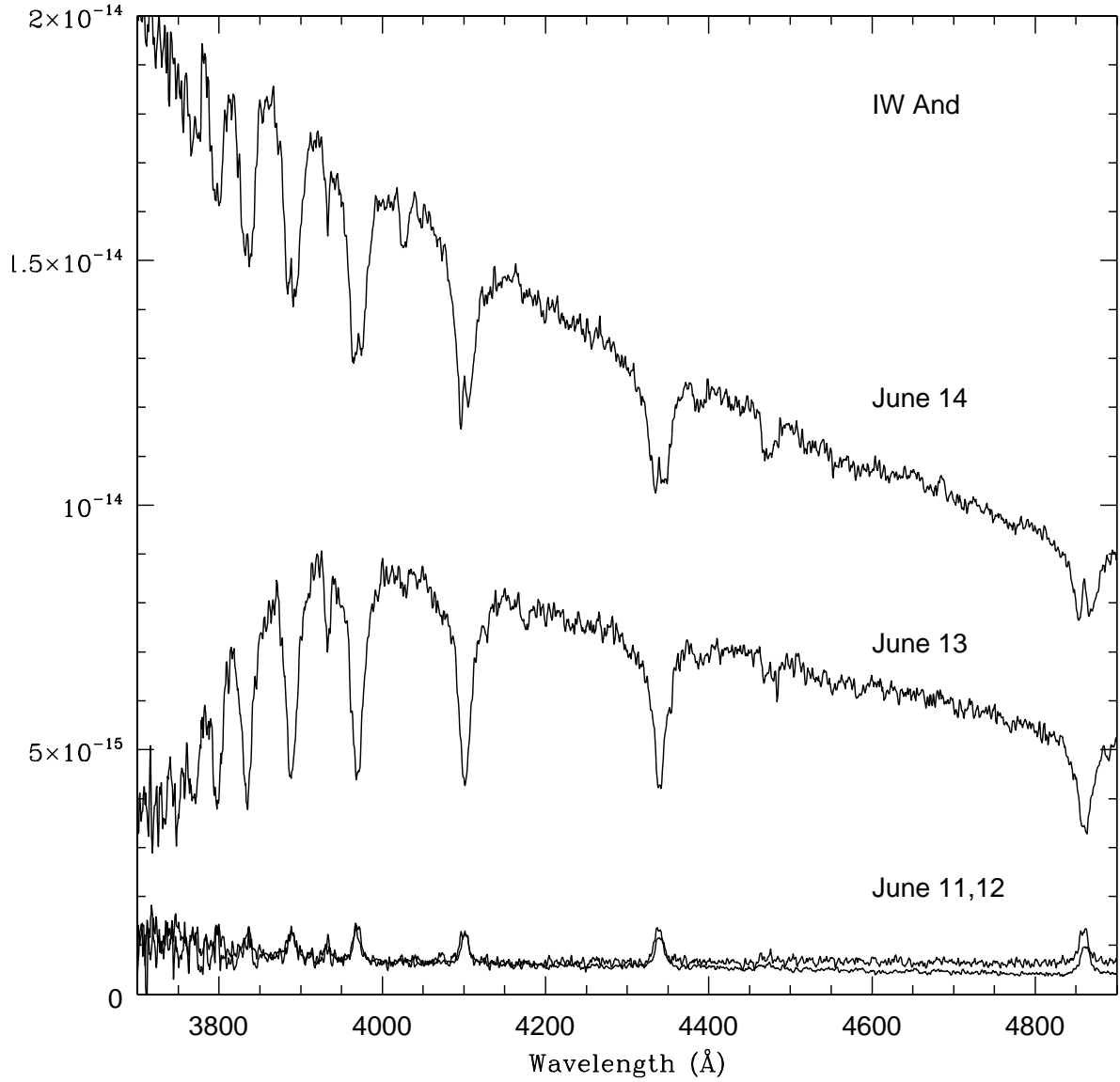}
\caption{KPNO sequence of spectra showing the change in spectra over
3 days when IW And went from quiescence to outburst.}
\end{figure}

\clearpage
\begin{figure}
\plotone{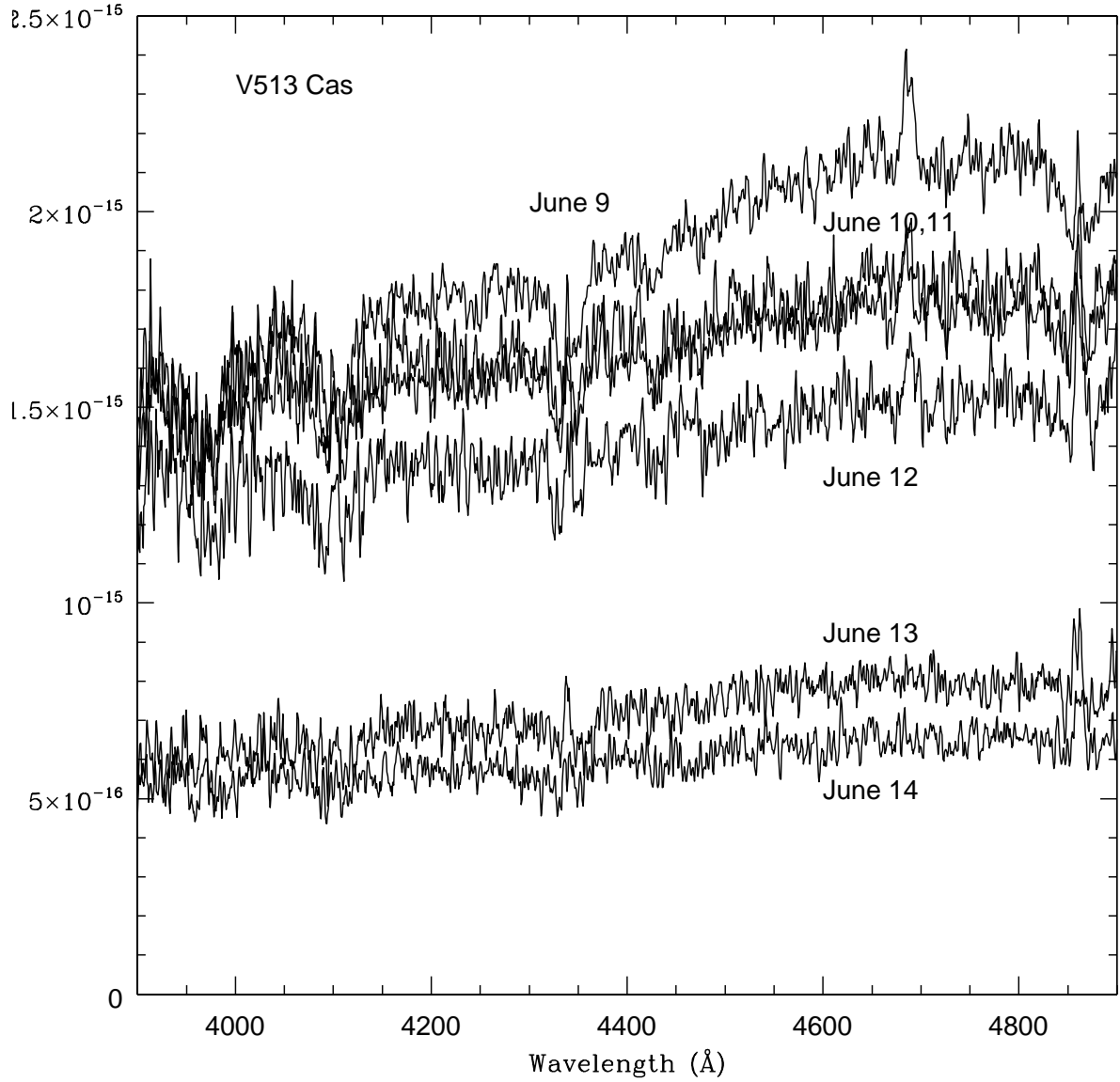}
\caption{KPNO sequence of spectra showing the change in spectra over
5 days as V513 Cas declined from outburst to midway to quiescence.}
\end{figure}

\clearpage
\begin{figure}
\plotone{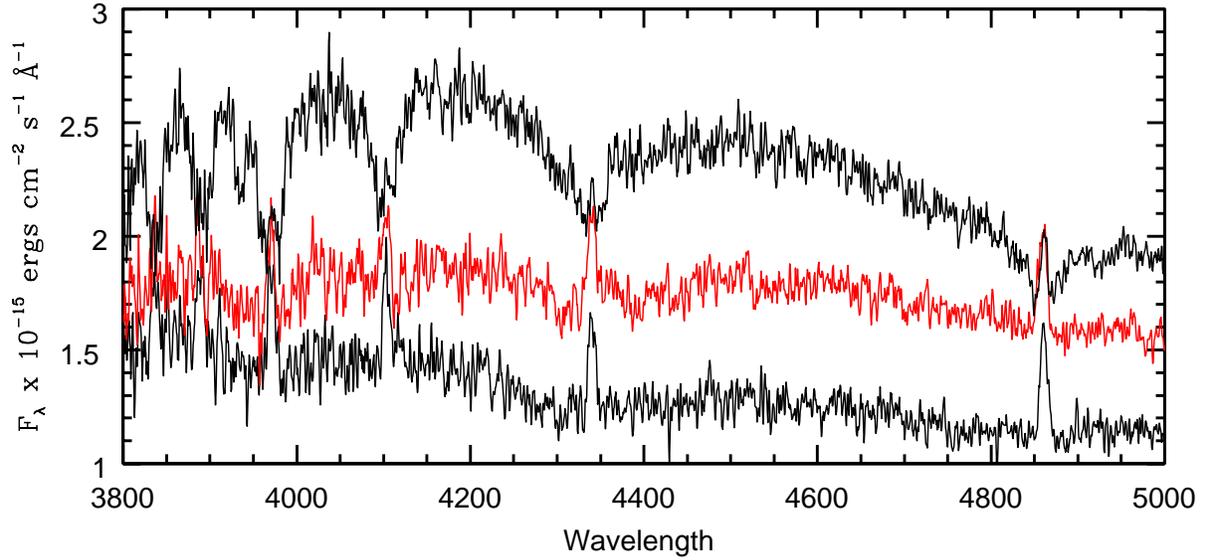}
\caption{APO sequence of spectra showing the change in spectra over
the last 2 hrs of 2012 October 29 as IW And changed from declining from
an outburst (bottom) to a standstill state (top).}
\end{figure}

\clearpage
\begin{figure}
\plotone{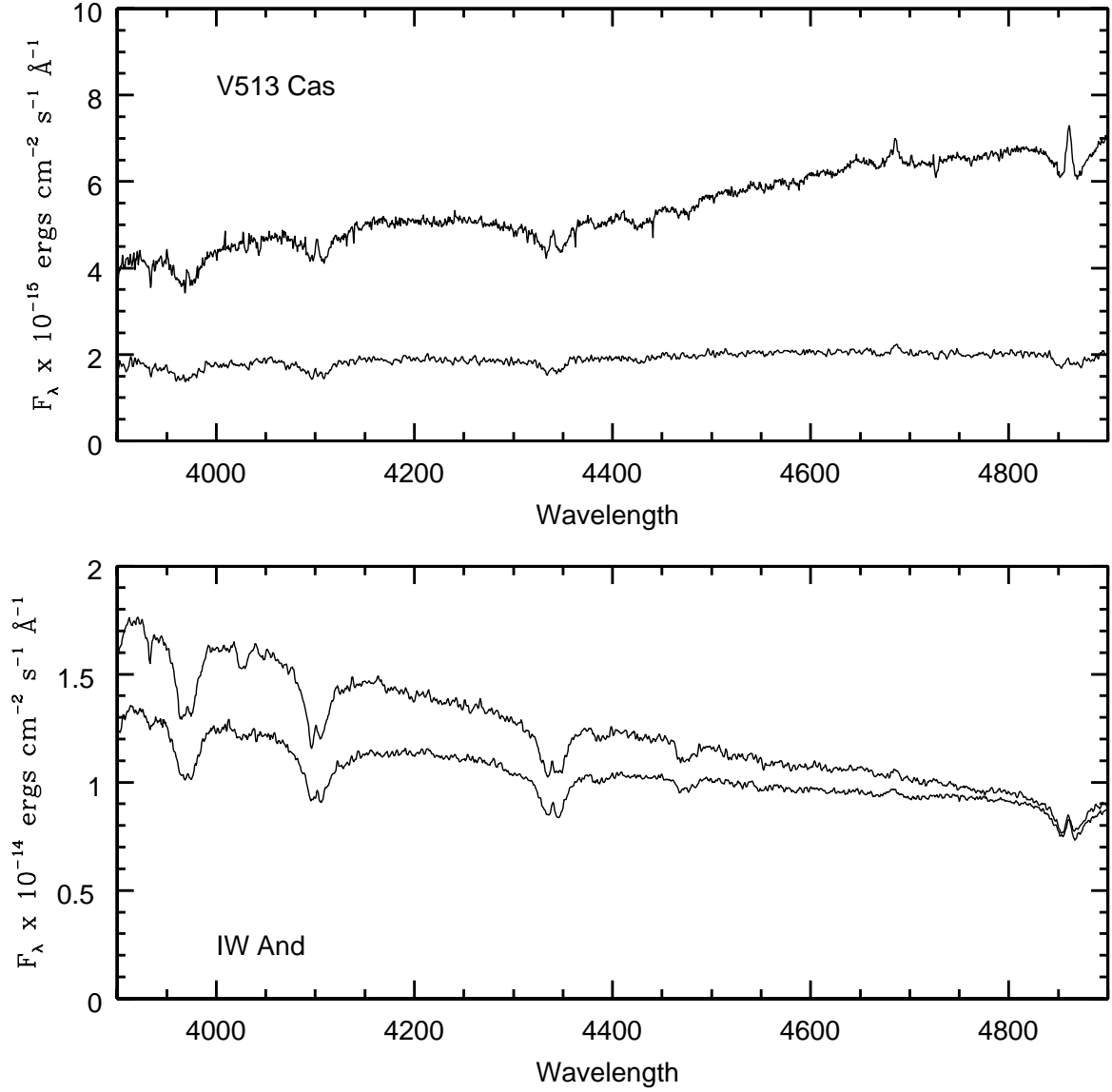}
\caption{Comparison of spectra at outburst and standstill for V513 Cas 
(2011 May 15 and 2012 October 6) 
and IW And (2011 June 14 and September 29).}
\end{figure}

\clearpage
\begin{figure}
\plotone{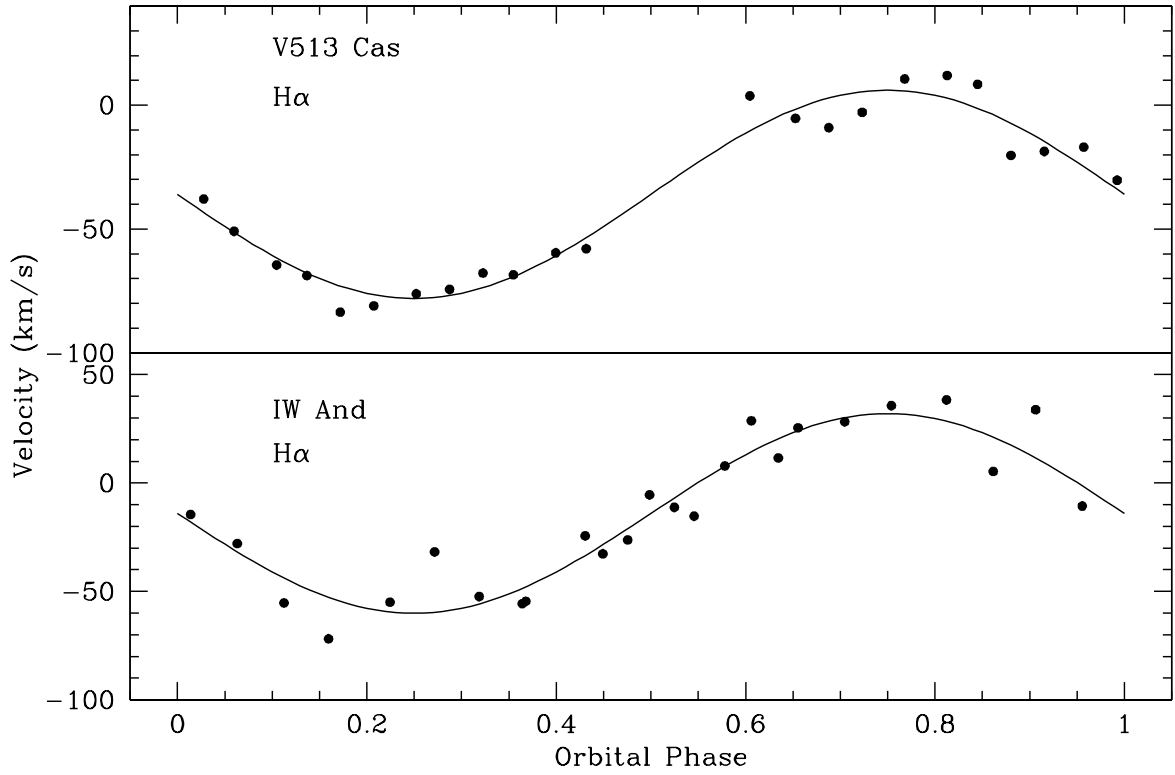}
\caption{H$\alpha$ velocities and best-fit sine-curve solution using
the values given in Table 3.}
\end{figure}

\clearpage
\begin{figure}
\plotone{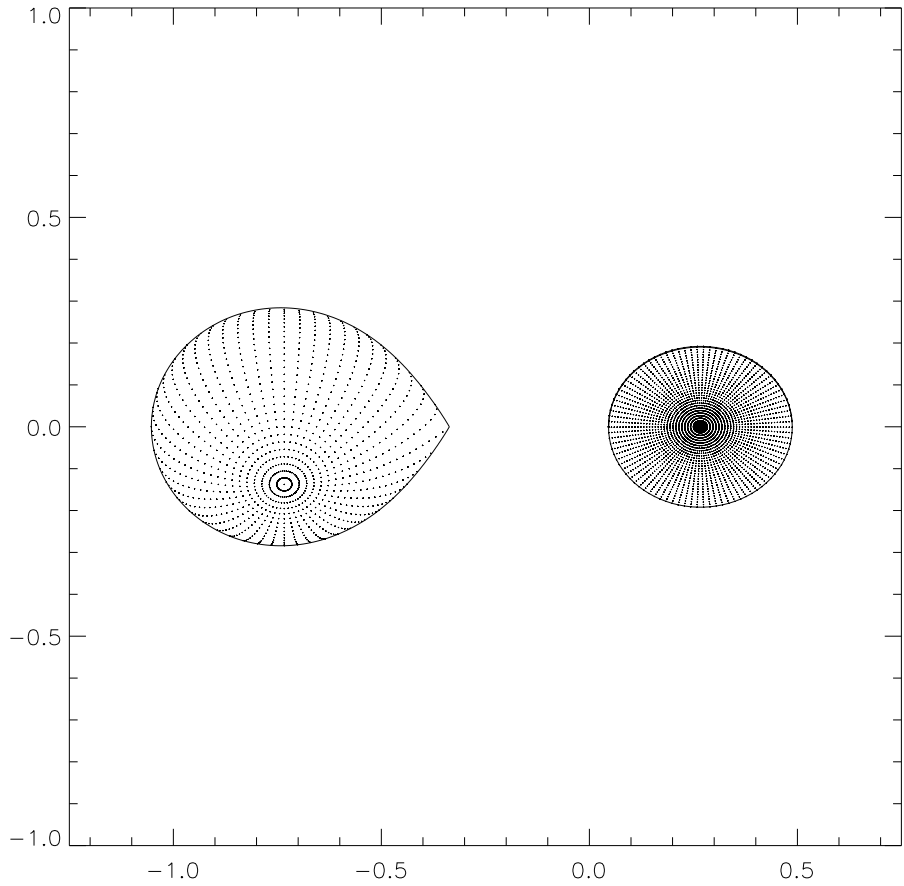}
\caption{View of IW And system model projected on the plane of the sky. The
orbital inclination is 30$^{\circ}$.}
\end{figure}

\clearpage
\begin{figure}
\plotone{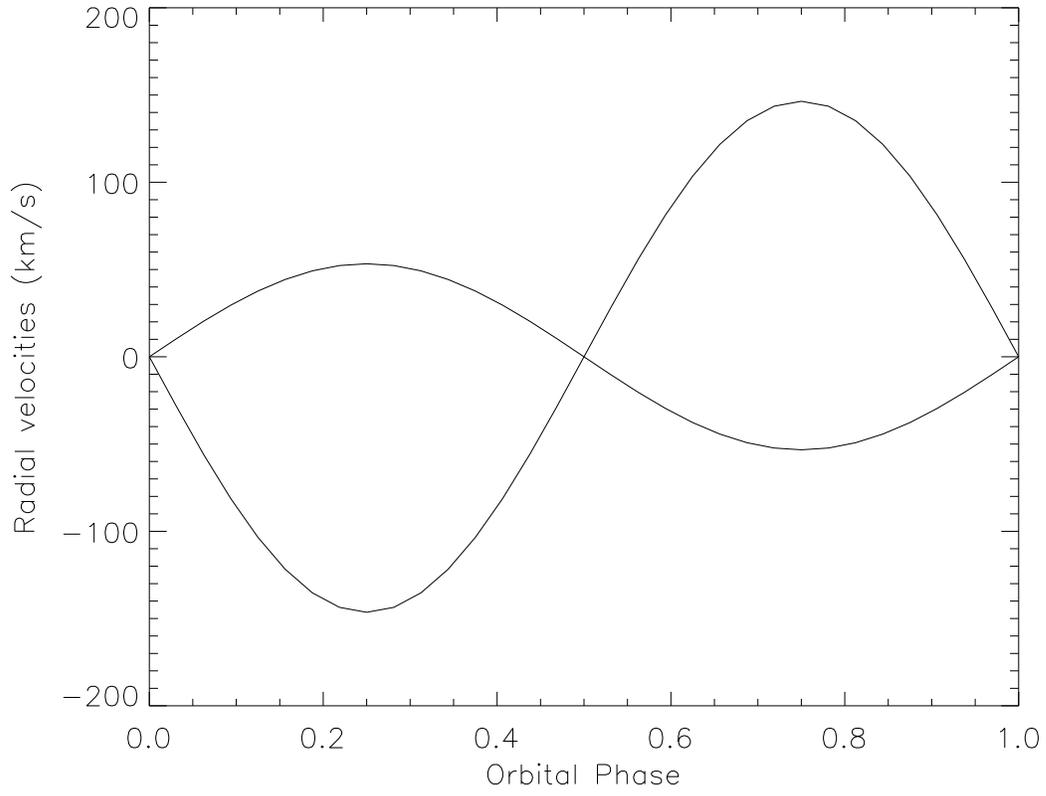}
\caption{Theoretical radial velocity curves of adopted model stellar system components.}
\end{figure}

\clearpage
\begin{figure}
\plotone{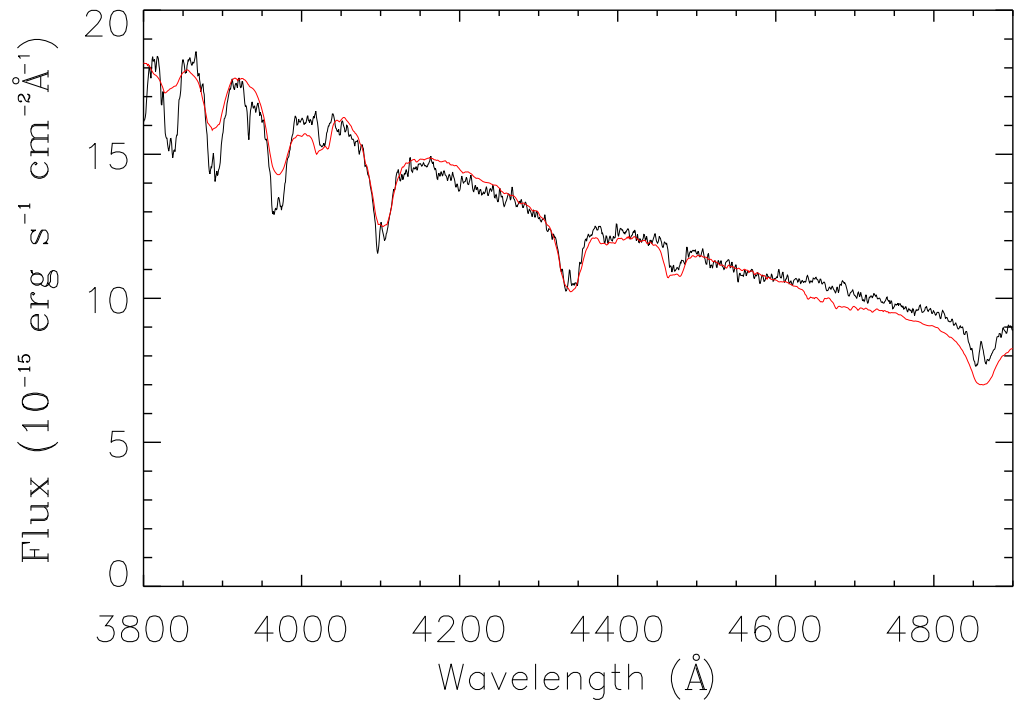}
\caption{Comparison of IW And outburst model (red curve) for mass accretion rate of
3$\times$10$^{-9}$M$_{\odot}$ yr$^{-1}$ with observed spectrum (black curve) from
2011 June 14.}
\end{figure}

\clearpage
\begin{figure}
\plotone{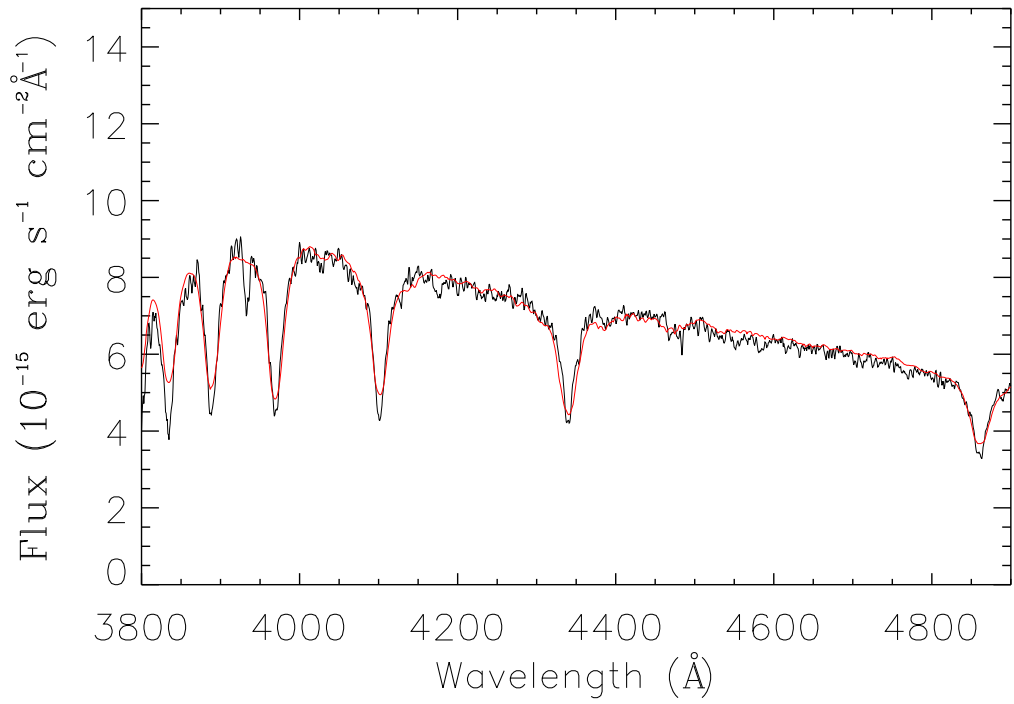}
\caption{Comparison of IW And rising to outburst model for mass accretion rate
of 3$\times$10$^{-9}$M$_{\odot}$ yr$^{-1}$ with observed spectrum from
2011 June 13.}
\end{figure}

\clearpage
\begin{figure}
\plotone{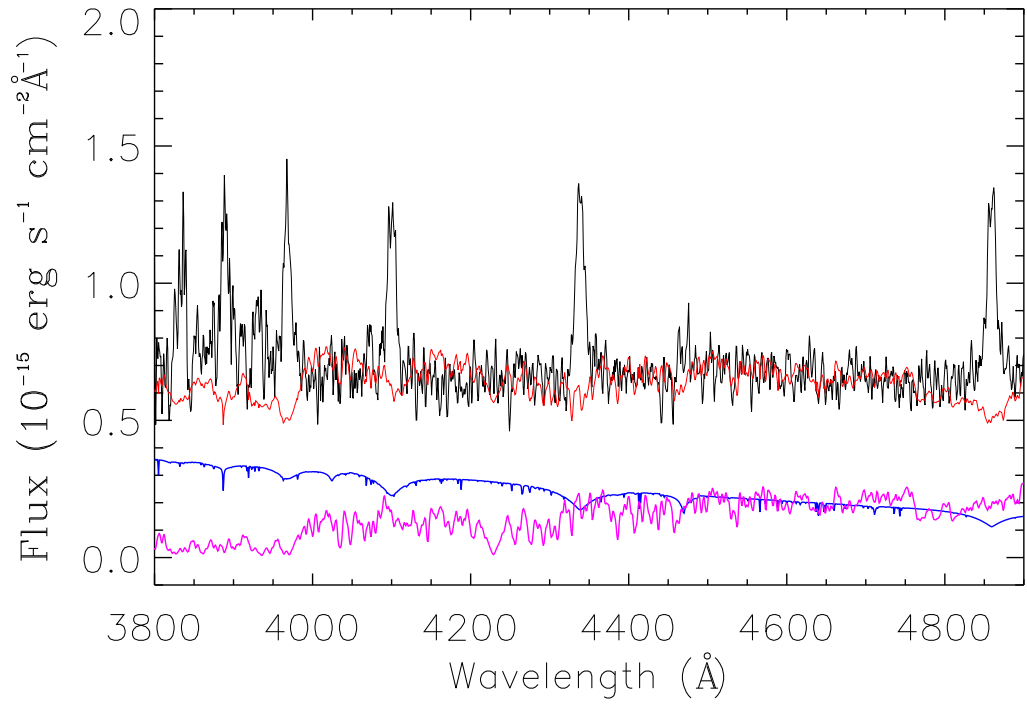}
\caption{Comparison of IW And quiescent model (red) for specified  T$_{eff}$ values with observed spectra
on 2011 June 11 (black). White dwarf component is shown with blue line
and secondary star with magenta line.}
\end{figure}

\clearpage
\begin{figure}
\plotone{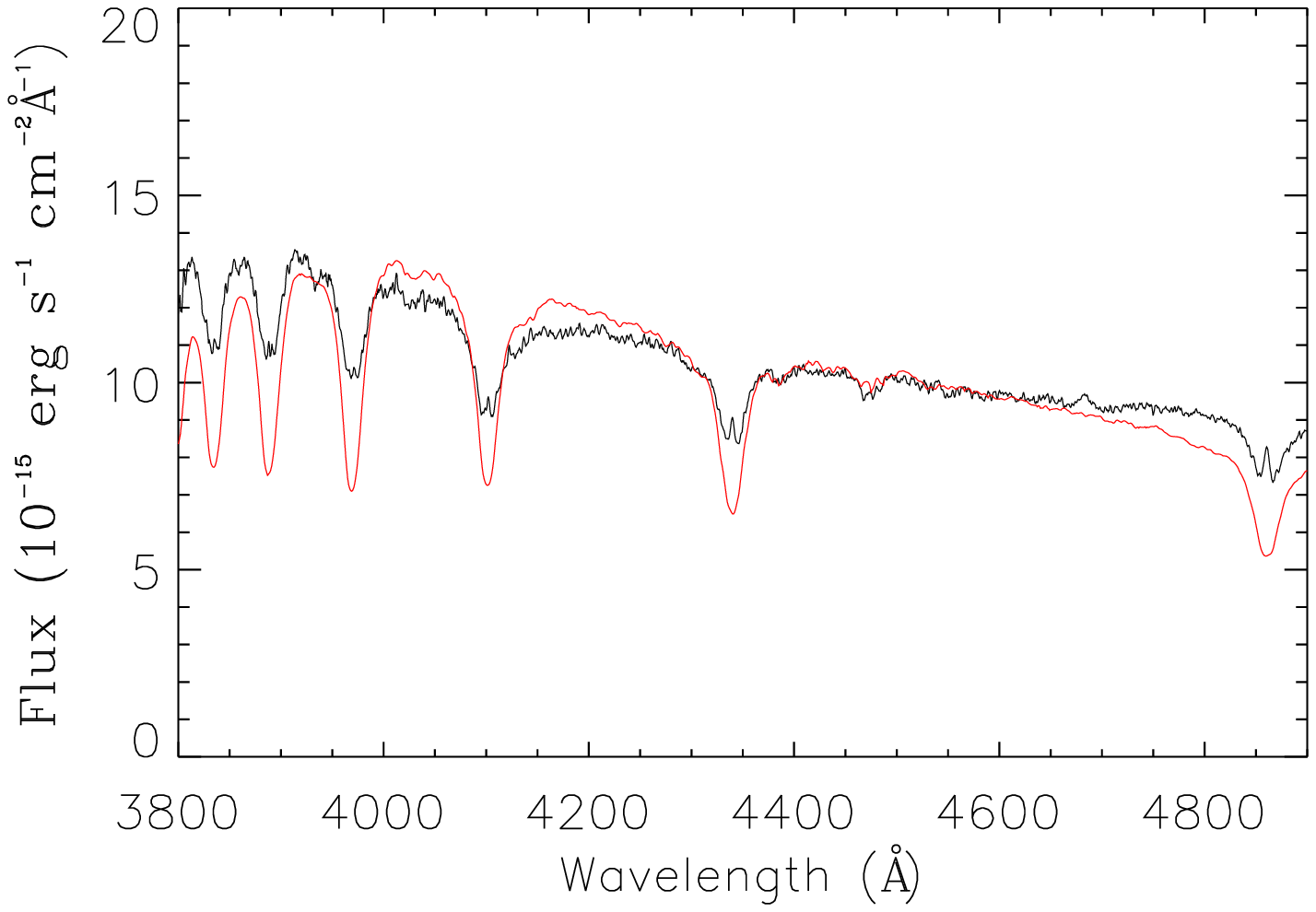}
\caption{Comparison of IW And standstill model for mass accretion rate
of 3$\times$10$^{-9}$M$_{\odot}$ yr$^{-1}$ with observed spectrum from
2011 September 29.}
\end{figure}

\clearpage
\begin{figure}
\plotone{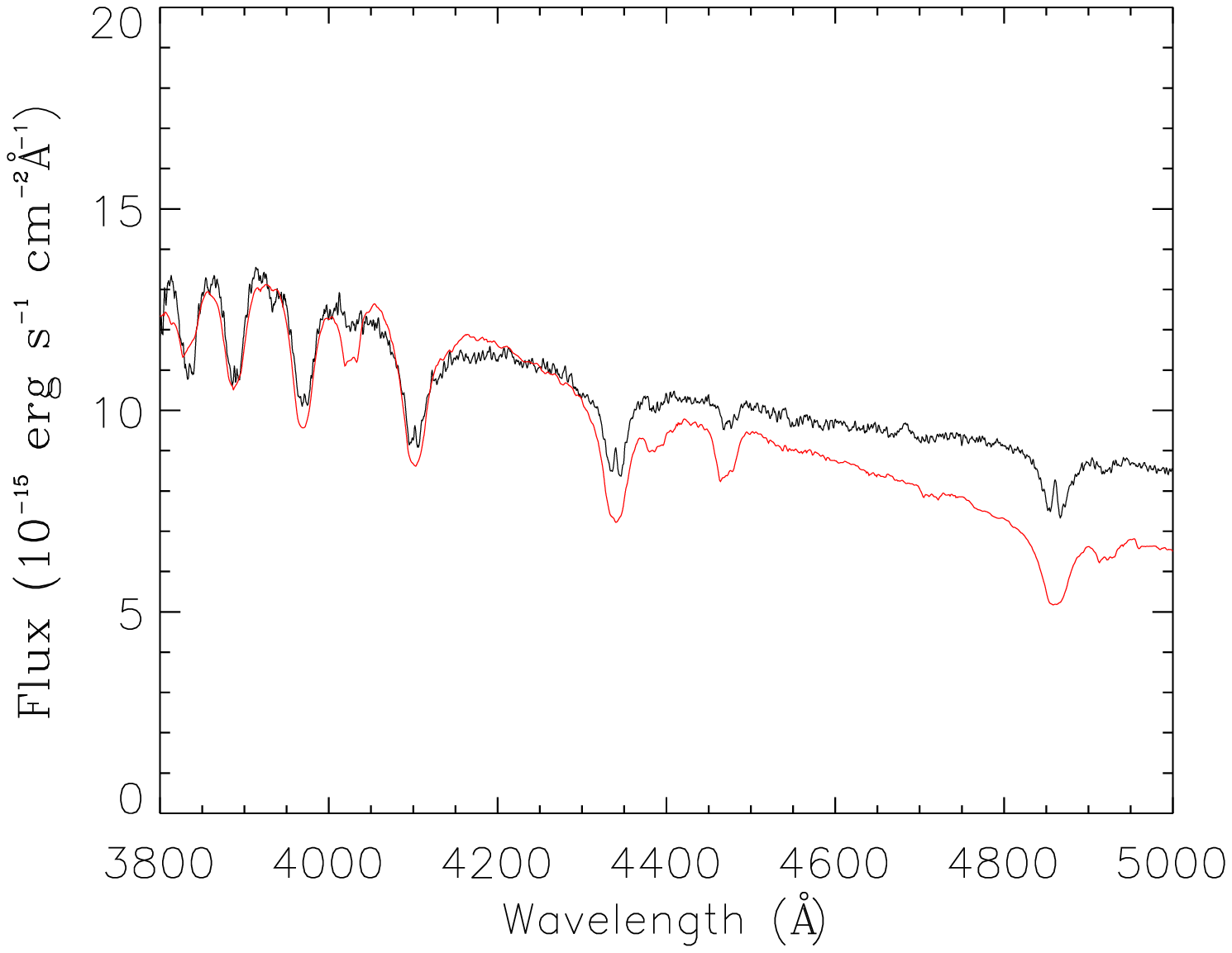}
\caption{Comparison of IW And standstill model for mass accretion rate
of 1$\times$10$^{-9}$M$_{\odot}$ yr$^{-1}$ with observed spectrum from
2011 September 29.}
\end{figure}

\end{document}